\documentclass[prl,twocolumn,amssymb,amsfonts,amsmath,superscriptaddress]{revtex4}
\usepackage{epsfig}

\begin{document}
\title{Geometry for the accelerating universe}

\author{Raffaele Punzi}
  \affiliation{Dipartimento di Fisica ``E. R. Caianiello'' Universit\`a di Salerno, 84081 Baronissi (SA) Italy, and INFN - Gruppo Collegato di Salerno, Italy}
\author{Frederic P. Schuller}\affiliation{Instituto de Ciencias Nucleares, Universidad Nacional Aut\'onoma de M\'exico, A. Postal 70-543, M\'exico D.F. 04510, M\'exico}
\author{Mattias N. R. Wohlfarth}
\affiliation{Center for Mathematical Physics and {II}. Institut f\"ur Theoretische Physik, Luruper Chaussee 149, 22761 Hamburg, Germany}
\begin{abstract} 
The Lorentzian spacetime metric is replaced by an area metric which naturally emerges as a generalized geometry in quantum string and gauge theory. Employing the area metric curvature scalar, the gravitational Einstein-Hilbert action is re-interpreted as dynamics for an area metric.  Without the need for dark energy or fine-tuning, area metric cosmology explains the observed small acceleration of the late Universe.
\end{abstract}
\maketitle
Assuming Einstein's equations for the gravitational field, explanations for the observed accelerating expansion of our Universe~\cite{Spergel,Knop} require some form of dark energy~\cite{copeland}. In its simplest incarnation, dark energy amounts to a positive cosmological constant, but its conceivable physical origins predict a value not even close to the required one~\cite{Weinberg:1988cp}. This has prompted attempts to deform the Einstein-Hilbert action for the spacetime metric~\cite{Carroll:2004de}, in order to avoid the introduction of dark energy. While partly successful, this typically just shifts the problem to the introduction of an unusually small deformation scale, and a largely ad hoc choice of the deformed action. The most serious stroke, however, is delivered by Lovelock's theorem~\cite{Lovelock:1972vz} which asserts that, in four dimensions, any action for a spacetime metric different from the Einstein-Hilbert action results in field equations of higher than second derivative order, with all the associated problems~\cite{Hindawi:1995an}.

\begin{figure}
\includegraphics[width=2.5in]{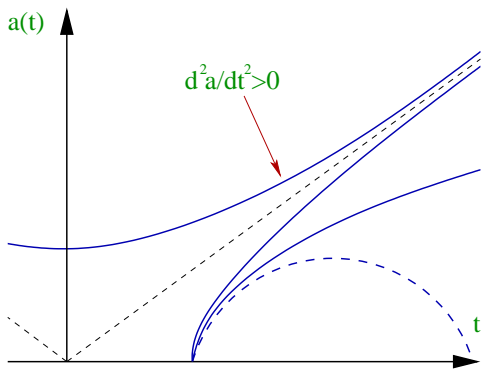}
\caption{\label{figure1}\emph{The solid curves sketch the string dust-filled area metric cosmologies: from bottom to top $\xi=0$, $\xi<0$ with $c>0$, and $\xi<0$ with $c<0$. The dashed curve depicts the inconsistent case $\xi>0$. The top curve is an important new result from area geometry (and cannot be obtained as a dust-filled Einstein cosmology): late-time acceleration tending to zero.}}
\end{figure}

In this Letter, we circumvent both Lovelock's theorem and the introduction of a deformation scale by reading the Einstein-Hilbert action not as an action for a spacetime metric, but as an action for an area metric. An area metric is a fourth rank tensor field, which measures two-dimensional tangent areas in close analogy to the way a metric measures tangent vectors~\cite{Schuller:2005ru}. We show that this yields a consistent classical gravity theory with second order field equations, which is a structural, and hence  rigid, extension of general relativity. A stunning prediction of area metric gravity is that the small accelerating expansion of our Universe at late times appears as a new exact solution of the field equations, without any additional assumptions such as dark energy or any form of fine-tuning, see figure~\ref{figure1}.

Far from being an exotic structure, area metrics naturally emerge in the quantization of fundamental theories whose classical formulation is based on a metric spacetime structure. This applies to gauge theories, gravity, and string theory: back-reacting photons in quantum electrodynamics effectively propagate in an area metric background~\cite{Drummond:1979pp}; canonical quantization of gravity \`a la Ashtekar~\cite{Ashtekar:1986yd} naturally leads to an area operator~\cite{Rovelli:1994ge}; the massless states of quantum string theory give rise to the Neveu-Schwarz two-form and dilaton besides the graviton, producing a generalized geometry which may be neatly absorbed into an area metric; the low energy action for D-branes~\cite{Fradkin:1985qd,Leigh:1989jq} is a true area metric volume integral~\cite{Schuller:2005ru}. But even for classical electrodynamics, the historical birthplace of metric spacetime, it has been observed early on that not a spacetime metric, but an area metric, presents the natural and most general background structure~\cite{Peres:1962}; then the deflection of light in gravitational fields becomes as rich as that in known optical media. In the mathematical literature, the idea to base geometry on an area measure goes back to Cartan~\cite{Cartan:1933}, who demonstrated that metric and area metric geometry are equivalent in three dimensions. This reconciles the postulate of an area metric spacetime with the fact that we can measure lengths and angles in three-dimensional spatial sections. 

In four dimensions, area metric geometry is a generalization of metric geometry: while every metric induces an area metric, not every area metric comes from an underlying metric. As mentioned above, the additional degrees of freedom may be viewed as arising from string theory, where generalized geometries play an increasingly important role: different ideas by Hitchin~\cite{Hitchin:2004ut} and Hull~\cite{Hull:2004in} have been applied to understand flux compactifications acted on with T-dualities~\cite{Kachru:2002sk} or mirror symmetry~\cite{Fidanza:2003zi}, and compactifications with duality twists~\cite{Dabholkar:2002sy}. The intimate relation of an area metric spacetime structure to string theory is further revealed by the fact that the minimal classical mechanical object one may discuss in an area metric background is a string. In the present note this plays a role in our discussion of fluids in area metric cosmology; these cannot consist of particles, but must be based on an integrable distribution of string worldsheets, which leads us to develop the notion of a string fluid. 

We now turn to a precise geometric formulation of the above ideas in four dimensions. Where detailed proofs are omitted, we refer the reader to our companion paper~\cite{companion} for full detail. Our guiding principle in the construction of area metric gravity is downward compatibility to metric spacetime, paying tribute to the phenomenal success of standard general relativity as a theory of gravity. So first consider a familiar metric manifold $(M,g)$ which naturally induces the area metric
\begin{equation}\label{Cg}
  C_{g\,abcd} = g_{ac}g_{bd} - g_{ad}g_{bd}\,,
\end{equation}
since $C_{g\,abcd}X^aY^bX^cY^d$ measures the squared area of a parallelogram spanned by vectors $X$ and $Y$. The basic idea of area metric geometry is to promote area metrics to a structure in their own right, by keeping salient algebraic properties of the metric-induced case: we define an area metric spacetime $(M,G)$ as a smooth four-dimensional manifold $M$ equipped with a fourth-rank covariant tensor field $G$ with the symmetries
\begin{equation}
G_{abcd}=G_{cdab}\,,\qquad G_{abcd}=-G_{bacd}\,,
\end{equation}
and the property that $G$ is invertible. Here invertibility is understood as follows: due to its symmetries the indices of $G$ may be combined to antisymmetric Petrov pairs $[ab]$ so that $G$ can be represented by a symmetric $6\times 6$ matrix which is required to be non-degenerate. The determinant $\mathrm{Det}\,G$ over this $6\times 6$ matrix gives rise to a volume form $\omega_G$ with components $\omega_{G\,abcd} = |\mathrm{Det}\,G|^{1/6} \epsilon_{abcd}$, where $\epsilon$ is the Levi-Civita tensor density normalized such that ${\epsilon_{0123}=1}$. The inverse area metric $G$ decomposes uniquely into a cyclic part $C^{a[bcd]}=0$ and a totally antisymmetric four-tensor that is dual to a scalar,
\begin{equation}\label{C+F}
  G^{abcd} = C^{abcd} + \phi\omega_C^{abcd}\,.
\end{equation}

We now construct area metric curvature tensors, which are downward compatible to their metric counterparts, in two steps.
First, note that $(M,G)$ gives rise to an effective metric $g_G$ entirely determined by area metric data. Using the above decomposition, we define
\begin{equation}
g_G^{ab} = \left.\frac{1}{2}\frac{\partial^2}{\partial p_a\partial p_b}\right|_{p=d\phi}\left(\mathcal{G}^{ijkl}p_ip_jp_kp_l\right)^{1/2}\,  
\end{equation}
for the Fresnel tensor
\begin{equation}
\mathcal{G}^{ijkl}=-\frac{1}{24}\omega_{C\,abcd}\omega_{C\,pqrs}C^{abp(i}C^{j|cq|k}C^{l)drs}\,.
\end{equation}
This construction can be motivated in detail by considering wave propagation on area metric manifolds in the geometric optics limit~\cite{companion}, where wavevectors $k$ obey the quartic Fresnel equation $\mathcal{G}^{abcd}k_ak_bk_ck_d=0$, see~\cite{Hehl:2004yk,Hehl:2005xu}. The effective metric generally does not contain all information of the area metric, which follows from a simple counting of the independent components. In particular, the re-induced area measure $C_{g_G}$ generally does not agree with $G$ in dimensions greater than three. But if we consider almost metric area metrics for which $C=C_g$ for some metric $g$ in the decomposition (\ref{C+F}), so that 
\begin{equation}\label{almet}
G^{abcd}=g^{ac}g^{bd}-g^{ad}g^{bc} + \phi\omega_g^{abcd}\,,
\end{equation}
then the effective metric $g_G$ recovers $g$ up to a sign, showing downward compatibility at the level of the effective metric. An important example for almost metric spacetimes is area metric cosmology: imposing isotropy and homogeneity on a four-dimensional area metric, by requiring that $\mathcal{L}_K G=0$ for the symmetry-generating vector fields $K$, leaves us with an area metric of the form~(\ref{almet}), where $g$ takes the standard Robertson-Walker form with scale factor $a(t)$ and normalized spatial curvature $k$, and the additional scalar $\phi$ depends only on time.

The effective metric $g_G$ gives rise to the torsion-free Levi-Civita connection $\nabla^{LC}$, but this is not of direct relevance for the area metric manifold  $(M,G)$. The area metric is not covariantly constant under $\nabla^{LC}$, so that, for instance, areas are not preserved under parallel transport~\cite{Schuller:2005ru}. We have to amend the connection so that $\nabla G=0$, which is unique up to a symmetry condition, and leads to the following covariant derivative~\cite{companion} on antisymmetric tensors $\Omega$:
\begin{equation}\label{Xdef}
  (\nabla_f \Omega)^{ab} = (\nabla^{LC}_f \Omega)^{ab} + \frac{1}{2}X^{ab}{}_{cdf} \Omega^{cd}\,,
\end{equation}
with the tensor $X^{ab}{}_{cdf}=\frac{1}{4}G^{abij} \nabla^{LC}_f G_{ijcd}$. This construction provides a true area metric connection, without using data additional to $G$. This completes the second step in our construction of the area metric curvature tensor, whose definition is now standard: $\mathcal{R}_G(X,Y)\Omega=[\nabla_X,\nabla_Y]\Omega-\nabla_{[X,Y]}\Omega$, so that
\begin{eqnarray}
 \!\!\!\!\!\!\!\! \mathcal{R}_G^{ab}{}_{cdij} & = & 4\delta^{[a}_{[c}R^{b]}{}_{d]ij}+\Big(\nabla^{LC}_i X^{ab}{}_{cdj}\Big.\nonumber\\
& & \Big.+\frac{1}{2}X^{ab}{}_{sti}X^{st}{}_{cdj}-(i\leftrightarrow j)\Big).
\end{eqnarray}
Natural contraction yields the area metric Ricci tensor, and the unique area metric curvature scalar of linear order in $\mathcal{R}_G^{a_1a_2}{}_{b_1b_2ij}$, the area metric Ricci scalar:
\begin{equation}
  \mathcal{R}_G{\,}_{mn} = \mathcal{R}_G^{pq}{}_{pmqn}\,,\qquad \mathcal{R}_G = g_G^{mn}\,\mathcal{R}_G{}_{mn}\,.
\end{equation}

All of the above curvature tensors reduce to their metric counterparts if the area metric is almost metric (\ref{almet}). Thus area metric geometry allows us to devise a gravity theory different from Einstein's without modifying the form of the Einstein-Hilbert action; replacing all metric quantities by their area metric counterparts, the latter is re-interpreted as an action for an area
metric manifold:
\begin{equation}\label{totalaction}
 S_{grav}\, + \,S_m \, = \frac{1}{2\kappa}\int_M\omega_G\,\mathcal{R}_G \, + \, \int_M \mathcal{L}_m\,.
\end{equation}
A gravitational constant $\kappa$ appears, and $\mathcal{L}_m$ represents the  matter Lagrangian density. Importantly, the equations of motion are now derived by variation with respect to the inverse area metric $G$, which leads to a fourth rank tensor equation
\begin{equation}\label{generalform}
K_{abcd}=\kappa T_{abcd}\,,
\end{equation}
where we have, schematically, the gravitational variation $K=|\mathrm{Det}\,G|^{-1/6}\delta(\kappa S_{grav})/\delta G^{-1}$ and the energy momentum tensor $T=-|\mathrm{Det}\,G|^{-1/6}\delta S_m/\delta G^{-1}$. Diffeomorphism invariance of the action implies an area metric Bianchi identity for the gravitational part of the action, and a conservation law for the area metric energy-momentum tensor. See \cite{companion} for the full computation. 

For applications to cosmology, where the area metric takes the almost metric form (\ref{almet}), the full equations of motion simplify considerably. 
Lengthy algebra reveals that in the almost metric case the gravitational variation $K$ is induced from a symmetric two-tensor and a scalar. If the energy-momentum tensor happens to be likewise induced from a symmetric second rank tensor $T_{ab}$ and a scalar $T^{\phi}$ then the full field equations~(\ref{generalform}), for $\tilde\phi=(1+\phi^2)^{-1/2}$, reduce to
\begin{eqnarray}\label{vacuumGg}
\kappa T_{ab} & = & R_{ab}-\frac{1}{2}Rg_{ab}-\tilde\phi^{-1}\left(\nabla_a\partial_b\tilde\phi-g_{ab}\square\tilde\phi\right),\nonumber\\
\kappa T^{\phi} & = & -\tilde\phi\left(1-\tilde\phi^2\right)^{1/2}R\,.
\end{eqnarray}
In particular, these equations are valid in vacuo; suitable field redefinitions reveal conformal equivalence to Einstein gravity minimally coupled to a massless scalar field. So the original vacuum theory is causal. We also conclude that any vacuum solution $(M,g)$ of Einstein gravity is a vacuum solution of area metric gravity (setting $\tilde\phi=1$ or $\tilde\phi\rightarrow 0$ with appropriate conditions on the derivatives). Including matter, however, it is no longer true that all solutions of Einstein gravity lift to solutions of area metric gravity. The energy-momentum tensor for Maxwell electrodynamics on area metric spacetime, for instance, is not induced from a symmetric two-tensor and a scalar. In other words, not only is an area metric spacetime a possible background for electrodynamics, but the backreaction via the area metric Einstein-Hilbert action requires a truly area metric background. 

The most prominent example of matter which \emph{is} compatible with almost metric backgrounds arises in area metric cosmology. Recall that in general relativity perfect fluids are the most general matter consistently coupling to cosmology, due to the way symmetries restrict the energy momentum tensor. This is also the case in area metric geometry, which however does not admit fluids based on point particles in the first place; we have to resort to string fluids based on continuous distributions of worldsheets. A three-component string fluid with local tangent surfaces $\Omega_I=\partial_t\wedge v_I$ for three $g$-spacelike vectors~$v_I$ is required to isotropically fill the spatial sections. Then the general source tensor $T_{abcd}$ equals
\begin{equation}\label{gensfluid}
\frac{\tilde\rho+\tilde p}{4}\sum_I G_{abij}\Omega^{ij}_IG_{cdkl}\Omega^{kl}_I + \tilde p\,G_{abcd} + (\tilde\rho+\tilde q)G_{[abcd]}
\end{equation}
in terms of three time-dependent functions $\tilde\rho$, $\tilde p$ and $\tilde q$. This string fluid tensor is induced by a two-tensor and by a scalar, as is needed for the equations to take the simple form (\ref{vacuumGg}). Using a time/space split,
\begin{eqnarray}
T_{00} & = & 12\frac{\tilde\rho-\tilde q\phi^2}{1+\phi^2}\,,\qquad T^\phi=\frac{24\phi\tilde q}{1+\phi^2}\,,\nonumber\\
T_{\alpha\beta} & = & 4\frac{-\tilde\rho+2\tilde p+\left(2\tilde\rho+2\tilde p+3\tilde q\right)\phi^2}{1+\phi^2}g_{\alpha\beta}\,.
\end{eqnarray}

Careful evaluation of the equations (\ref{vacuumGg}) now reveals a remarkable correspondence between, on the one hand, area metric cosmology (determined by $g$ and $\phi$) plus string fluid matter ($\tilde\rho$, $\tilde p$ and $\tilde q$), and, on the other hand, Einstein cosmology ($g$) plus perfect fluid matter ($\rho$ and $p$). The effective energy density and pressure of the perfect fluid emerge as $\rho=x-y$ and $p=x+y$ for
\begin{equation}
x = -H\dot{\tilde\phi}\tilde\phi^{-1}+4\kappa\left(\tilde\rho+\tilde q\right)\tilde\phi^2\,,\qquad y=4\kappa\tilde q
\end{equation}
where $H$ is the Hubble function. Recall ${\tilde\phi=(1+\phi^2)^{-1/2}}$ which obeys $\square \tilde\phi=\partial V(\tilde\phi)/\partial\tilde\phi$ with potential
\begin{equation}\label{pote}
V(\tilde\phi)=4\kappa\left(\tilde\rho+\tilde p+\tilde q\right)\tilde\phi^2-4\kappa\left(\tilde\rho+\tilde q\right)\tilde\phi^4\,.
\end{equation} 
For the effective equation of state parameter $w=p/\rho$ one obtains the following results, generic for any string fluid and thus for general matter in area metric cosmology. The limit $y/x\rightarrow 0$ gives $w=1/3$, i.e., an effective radiation fluid. This is exactly realized by area metric vacuum cosmology ($y=0$). The limit $x/y\rightarrow 0$ gives $w=-1/3$, i.e., a universe with zero acceleration $\ddot a=0$. The condition $w<-1/3$ for an accelerating universe is satisfied if either $y<x<0$ or $y>x>0$. For values $y$ close to the pole in $w$ at $y=x$, one may obtain any value of $w$, both positive or negative. So string fluids in principle are able to describe any physical universe. 

For the late universe, matter has spread out so much that interactions are no longer important, so that we may specialize to non-interacting string dust, whose parameters $\tilde p$, $\tilde\rho$ and $\tilde q$ can be identified from the condition that energy conservation should be implied by the minimal surface equation and by a generalized continuity equation \cite{companion}, which yields
\begin{equation}\label{sdust}
\tilde p=0\,,\qquad \tilde q=-\tilde\rho\,.
\end{equation}
We now provide exact solutions for the string dust case with $\tilde\rho\neq 0$. Using $R_{00}=-3\dot H-3H^2$ and $R_{\alpha\beta}=(2ka^{-2}+\dot H+3H^2)g_{\alpha\beta}$ one collects the full set of equations from (\ref{vacuumGg}) and (\ref{pote}). The equations of area metric cosmology filled with string dust then become
\begin{equation}
\tilde\phi = \lambda \dot a\,,\qquad \tilde\rho = \zeta a^{-2}\,,\qquad 0=\ddot a+\frac{\dot a^2}{a}+\frac{k-4\kappa\zeta}{a}
\end{equation}
for integration constants $\lambda$ and $\zeta$, while the effective equation of state parameter for the perfect fluid created by area metric cosmology can be rewritten as
\begin{equation}\label{parst}
w=\frac{1}{3}-\frac{8\kappa}{3}\frac{\zeta}{\dot a^2+k}\,.
\end{equation}
The equation for $a$ is exactly solved by the scale factor
\begin{equation}
a(t)=\left\{\begin{array}{cl}

\sqrt{c\left(t-t_0\right)} & \textrm{for }\xi=0\,,\\
\sqrt{c\xi^{-1}-\xi\left(t-t_0\right)^2} & \textrm{for }\xi\neq 0\,,

\end{array}
\right.
\end{equation}
for integration constants $c$ and $t_0$, and $\xi=k-4\kappa\zeta$. We now discuss the possible cases for $\zeta>0$ ($\zeta<0$ is similar, but does not allow consistent $k=0$ cosmologies).

It turns out that the case $\xi>0$ does not provide a consistent late-time solution, and the case $\xi=0$ only does so for one particular value of $\zeta$, which is physically excluded by fine-tuning arguments. For details see \cite{companion}. This leaves us with the truly interesting case $\xi<0$. Here $k<4\kappa\zeta$, and so this can be realized by $k=-1$ and $k=0$ cosmologies without further restrictions, but also by $k=+1$ cosmologies with matter density $4\zeta>1/\kappa$. The effective equation of state parameter takes the form\begin{equation}\label{parg0}
w=\frac{1}{3}-\frac{8\kappa\zeta\left(c-\xi^2\left(t-t_0\right)^2\right)}{3kc-12\kappa\zeta\xi^2\left(t-t_0\right)^2}\,.
\end{equation}
We have to discuss two subcases. If $c>0$ then the solution is defined for $t-t_0>\sqrt{c/\xi^2}$. If $c<0$ then the solution is defined for all $t$. For both signs of $c$ we obtain the late time limit of $w\rightarrow -1/3$. We can also check consistency for late times $t-t_0\rightarrow\infty$: if $0<\lambda\sqrt{-\xi}$ this limit ensures $0\leq\tilde\phi<1$. The solution with $c>0$ corresponds to an open, eternally decelerating, universe with initial singularity, for which the acceleration tends to zero for late times. The solution for $c<0$ describes an open, eternally \emph{accelerating}, universe without singularities that passes a minimal radius $a(t_0)=\sqrt{c\xi}$ and has a late-time acceleration tending to zero, see figure \ref{figure1}.

So we have shown that non-interacting string dust matter in area metric cosmology provides a more interesting phenomenology than Einstein cosmology filled with point particle dust. Most strikingly, area metric cosmology naturally explains, for all $k=0$ and $k=\pm 1$ cosmologies, the late-time acceleration of the universe! This explanation neither invokes concepts such as a cosmological constant or dark energy, nor does it invoke fine-tuning since the acceleration automatically tends to small values at late time; these results become consequences of area metric geometry applied to the very simplest scenario, cosmology. It is remarkable that area metric geometry at the same time emerges from standard physical theories in a number of different places, and also, without further assumptions, addresses one of the most intriguing cosmological observations of recent times.

\acknowledgments{We have the pleasure to acknowledge useful comments by Gary Gibbons, Julius Wess, Alexander Turbiner, Albert Schwarz, Friedrich Hehl and Marcus Werner. This work has been supported by the European Union (RP), DGAPA-UNAM grant IN121306 (FPS) and the DFG Emmy Noether grant WO 1447/1-1 (MNRW).}


\end{document}